# A Review of New CMOS Material Platforms for Integrated Nonlinear Optics

**Roberto Morandotti**
*INRS-EMT, 1650 Boulevard Lionel Boulet, Varennes, Québec, Canada, J3X 1S2*

**David J. Moss**
*School of Electrical and Computer Engineering (SECE)*
*RMIT University, Melbourne, Australia 3001*

**Alexander L. Gaeta**
*School of Applied and Engineering Physics, Cornell University,*
*Ithaca, New York 14853, USA*

**Michal Lipson**
*School of Electrical and Computer Engineering, Cornell University,*
*Ithaca, New York 14853, USA*

**Nonlinear photonic chips have enabled the generation and processing of signals using only light, with performance far superior to that possible electronically - particularly with respect to speed. Although silicon-on-insulator has been the leading platform for nonlinear optics, its high two-photon absorption at telecommunications wavelengths poses a fundamental limitation. We review recent progress in non-silicon CMOS-compatible platforms for nonlinear optics, with a focus on $Si_3N_4$ and Hydex. These material systems have opened up many new capabilities such as on-chip optical frequency comb generation and ultrafast optical pulse generation and measurement. This review highlights their potential impact as well as the challenges to achieving practical solutions for many key applications.**

# Introduction

All-optical signal generation and processing [1,2] have been highly successful at enabling a vast array of capabilities, such as switching and de-multiplexing of signals at unprecedented speeds [3,4], parametric gain [5] on a chip, Raman lasing [6], wavelength conversion [7], optical logic [8], all-optical regeneration [9,10], radio-frequency (RF) spectrometry at THz speeds [11,12], as well as entirely new functions such as ultra-short pulse measurement [13,14] and generation [15] on a chip, optical temporal cloaking [16], and many others. Phase sensitive functions [14, 17], in particular, will likely be critical for telecommunications systems that are already using phase encoding schemes [18,19]. The ability to produce these devices in integrated form - photonic integrated circuits (PICs) - will reap the greatest dividends in terms of cost, footprint, energy consumption and performance, where demands for greater bandwidth, network flexibility, low energy consumption and cost must all be met.

The quest for high performance platforms for integrated nonlinear optics has naturally focused on materials with extremely high nonlinearity, and silicon-on-insulator (SOI) has led this field for several years [2]. Its high refractive index allows for tight confinement of light within SOI nanowires that, when combined with its high Kerr nonlinearity ($n_2$), yields an extremely high nonlinear parameter $\gamma = 300,000$ $W^{-1}$ $km^{-1}$ ($\gamma = \omega\, n_2 / c\, A_{eff}$, where $A_{eff}$ is the effective area of the waveguide, $c$ is the speed of light and $\omega$ is the pump frequency).

As a platform for linear photonics, SOI has already gained a foothold as a foundation for the silicon photonic chip industry [20] aimed at replacing the large interface cards in optical communications networks, due to its ability to combine electronics and photonics on the same chip. These first-generation silicon photonic chips typically employ photonics for realizing passive splitters, filters, and multiplexers, and offer the substantial benefits of exploiting the broadly established global CMOS infrastructure, although not without challenges [21, 22]. What is clear, however, is that any progress made in this direction for these first generation SOI chips will have a direct benefit on future generation all-optical chips, whether in SOI directly or in other CMOS-compatible platforms. Many of the issues that make CMOS compatibility compelling for linear passive devices apply equally well to all-optical-processing chips.

Over the last 10 years the achievements reported in SOI fabricated using CMOS technology that exploit both linear and nonlinear optical phenomena [2, 23] are impressive [23 - 39], ranging from optical buffers [24], optical interconnects [25-27], ring resonators [28, 29], Raman gain and lasing [30 - 32], time lensing [13, 33], slow light based on photonic crystals [34 - 36], optical regeneration [9], parametric gain [5, 37 - 39] to the promise of direct optical transitions [40, 41], and even correlated photon pair generation [42].

Indeed, SOI is so attractive as a platform for both linear *and* nonlinear photonics that, were it not for one single issue, the quest for the ideal all-optical platform might already be largely solved. That issue is the fact that bulk crystalline silicon suffers from high nonlinear absorption due to two-photon absorption (TPA) [23, 43] in all telecommunications bands at wavelengths shorter than about 2000 nm. While the effect of TPA generated free carriers can be mitigated by the use of p-i-n junctions to sweep out carriers [44], for example, silicon's *intrinsic* nonlinear figure of merit (FOM = $n_2/\beta \lambda$, where $\beta$ is the TPA coefficient and $\lambda$ the wavelength) is only 0.3 near 1550 nm [45 - 47]. This represents a *fundamental* limitation – being an intrinsic property of silicon's bandstructure. It cannot be compensated for by, for example, engineering waveguide dimensions etc.. The fact that many impressive all-optical demonstrations have been made in silicon despite its low FOM is a testament to how exceptional its linear and nonlinear optical properties are. Nonetheless, the critical impact of silicon's large TPA was illustrated [48, 49] in 2010 by the demonstration of high parametric gain at wavelengths beyond 2 μm, where TPA vanishes. Indeed, it is likely that in the mid-infrared wavelength range where it is transparent to both one photon and two photon transitions –between 2 and 6 μm – silicon will undoubtedly remain a highly attractive platform for nonlinear photonics.

For the telecom band, however, the search has continued for the ideal nonlinear platform. Historically, two important platforms have been chalcogenide glasses (ChG) [50] and AlGaAs [51]. Chalcogenide glasses have achieved very high performance as nonlinear devices, but realising nanowires with ultrahigh nonlinearity (> 10,000 $W^{-1}km^{-1}$) has proven elusive due to fabrication challenges, as has also been the goal of achieving a material reliability comparable to that of semiconductors. AlGaAs was the first platform proposed for nonlinear optics in the telecom band [51] and offers the powerful ability to tune the nonlinearity and FOM by varying the alloy composition. A significant issue for AlGaAs, however, is that nanowires require challenging fabrication [52] methods – particularly etching – in order to realize very high, narrow mesa structures. Nonetheless, both platforms offer significant advantages and will undoubtedly play a role in future all-optical photonic chips.

Another promising approach has involved the use of hybrid integration of SOI nanowires – usually in the form of a "slot" cross-section – with nonlinear organic polymers [53, 54]. This approach has the advantage of exploiting the SOI platform for its ability for extremely efficient mode confinement, while at the same time forcing a high proportion of the mode field into slot filled with polymer material, thus exploiting the high nonlinearity and FOM of the polymer.

This review focuses on new platforms recently introduced [55-59] for nonlinear optics that have achieved considerable success and that also offer CMOS compatibility. They are based on silicon nitride ($Si_3N_4$) and high-index doped silica (trade-named

Hydex). Originally developed for linear optics [60 - 62], these platforms are particularly promising due to their low linear loss, relatively large nonlinearities compared to typical fibers and, most significantly, their negligible nonlinear loss at telecommunication wavelengths [63]. In addition, their high quality CMOS-compatible fabrication processes, high material stability, and the ability to engineer dispersion [57] make these platforms highly attractive.

Indeed, within a short period of time significant progress has been made with respect to their nonlinear performance - particularly in the context of optical frequency comb (OFC) generation in microresonators [64 - 71]. Since the demonstration of OFCs in SiN [57] and Hydex [58] in 2010, this field has proliferated. Extremely wide-band frequency combs [64 - 66], sub 100-GHz combs [67], line-by-line arbitrary optical waveform generation [68], ultrashort pulse generation [15, 70], and dual frequency combs [71] have been reported. In addition, OFC harmonic generation [72] has been observed. These breakthroughs have not been possible in SOI at telecom wavelengths because of its low FOM.

Here, we review the substantial progress made towards nonlinear optical applications of these CMOS-compatible platforms and briefly discuss the newly emerging promising platform of amorphous silicon. This review is organized according to application, beginning with basic nonlinear optics including four wave mixing, parametric gain etc., followed by applications to on-chip optical sources (frequency combs, OPOs etc.) and finally applications to optical pulse measurement. The high performance, reliability, manufacturability of all of these platforms combined with their intrinsic compatibility with electronic-chip manufacturing (CMOS) has raised the prospect of achieving practical platforms for future low-cost nonlinear all-optical PICs.

## Nonlinear Optics in Nanowires and Waveguides

### $Si_3N_4$ Nanowires

Silicon nitride ($Si_3N_4$), a CMOS-compatible material well known to the computer chip industry as a dielectric insulator, has been used as a platform for linear integrated optics [60] for some time. However, only recently [55] it has been proposed as a platform for nonlinear optics. Historically, the challenge for SiN optical devices has been to grow low loss layers thicker than 250 nm, due to tensile film stress. Achieving such thick layers is critical for nonlinear optics since both high mode confinement as well as dispersion engineering for phase matching [57] are needed. Thick (>500 nm) low loss SiN layers were recently grown (Figure 1) by plasma-enhanced chemical vapor deposition (PECVD) [55] as well as by low-pressure chemical vapor deposition (LPCVD) [57]. The latter approach employed a thermal cycling process that resulted in very thick (700 nm) films that yielded nanowires with very low propagation loss (0.4 dB/cm).

The first nonlinear optical studies of SiN waveguides were reported in 2008 [55] (Fig. 1a), showing nonlinear shifting of the resonances in 700 nm thick SiN ring resonators obtained with 200 mW CW optical pump power. Time resolved measurements enabled thermal and Kerr contributions to be separated, resulting in an $n_2 \sim 10\times$ that of silica glass, which is consistent with Millers rule [73]. This value has been validated in subsequent reports of nonlinear optics in SiN nanowires and resonators.

Parametric gain in SiN was first demonstrated [57] in low loss nanowires by centering the pump for the FWM process in the anomalous group-velocity dispersion (GVD) regime near the zero-GVD point. This allowed for broad-bandwidth phase matching, and hence signal amplification, over a wide range of wavelengths. Net gain was achieved in long (6 cm) SiN waveguides with a nonlinear $\gamma$ parameter of 1,200 $W^{-1}km^{-1}$ and a zero-GVD point near 1,560 nm. An on/off signal gain as high as 3.6 dB was observed over a 150-nm bandwidth, and since the total propagation loss through the waveguides was 3 dB, this represented a net parametric amplification.

## Hydex Waveguides

Hydex glass was developed [62] as a low loss CMOS compatible optical platform primarily for the realization of advanced linear filters. Its refractive index range of $n = 1.5$ to 1.9 is slightly lower than that of SiN, being comparable to SiON, and so a buried waveguide geometry is typically used rather than nanowires. Nonetheless, the core-cladding contrast of 17% still allows for relatively tight bend radii of 20 μm. The proprietary nature of the composition of Hydex is solely aimed at reducing the need for high temperature annealing by mitigating the effect of N-H bonds - the main source of absorption loss in the telecom band. This means that as-grown films intrinsically have low loss, making the growth process more compatible with CMOS processes.

The success of the Hydex platform in nonlinear optics can be primarily attributed to its extraordinarily low losses – both linear and nonlinear. Linear propagation losses as low as 5 - 7 dB/meter have been achieved, allowing for the use of extremely long waveguide spirals [14, 63]. Figure 1b shows a schematic of a 45-cm-long spiral waveguide contained within a square area of 2.5 mm x 2.5 mm and pigtailed to a single-mode fiber via low loss on-chip beam expanders, as well as a SEM picture of its cross section (before cladding deposition). The films were fabricated with CMOS compatible processes that yielded exceptionally low sidewall roughness in the core layer.

For Hydex, self-phase modulation experiments [63] yielded a Kerr nonlinearity of $n_2 = 1.15 \times 10^{-19}$ $m^2/W$, or ~ 4.6 × silica glass, and roughly half as large as that for SiN, with a nonlinearity parameter $\gamma \cong 233 W^{-1}km^{-1}$ (~ 200 × standard single-mode telecommunications fibers). Like $Si_3N_4$, this enhancement in $n_2$ is in agreement with Miller's rule [73]. This implies that the enhancement in $n_2$ arises solely from the increase in the linear refractive index and not from the proprietary aspect of the material composition.

Although self-phase modulation measurements are not overly dependent on dispersion (for waveguide lengths small compared with the dispersion length), for four-wave mixing, including parametric gain, dispersion is critical for efficient, wide bandwidth FWM including parametric gain [74, 75]. Hydex waveguides have been engineered to yield anomalous dispersion over most of the C-band with the zero-GVD points being 1600 nm for a TE polarization and 1560 nm for TM [76]. This resulted in a large FWM wavelength tuning range with an efficient parametric gain of +15dB and a signal to idler conversion efficiency of +16.5dB [74].

Finally, we note that the broad use of the description "CMOS compatible" in this context is intended to reflect a general compatibility in terms of growth temperatures (< 400 C°) and materials that are used in the CMOS process (silicon nitride, Hydex, and silicon oxynitride). It does not address the complexities and challenges of integrating optical and electronic devices with substantially different size scales; nor does it address the challenges of adapting CMOS production lines to optical device fabrication, both of which have been discussed at length [21, 22]. A central problem in terms of integrating waveguides and nanowires with electronic components are the relatively large thicknesses of both the core and cladding films. In this regard, the higher refractive index contrast of SiN (about 0.5) compared to Hydex (about 0.3) is a significant advantage. Both materials, however, require considerably thicker core and cladding layers than SOI - this is probably a key area in which SOI out-performs these platforms. Nonetheless, the concept of CMOS compatibility presented here, which these new platforms satisfy, is a powerful one that will go a long way towards enabling the broad application of CMOS techniques and manufacturing infrastructure to nonlinear photonic chips.

## Microresonator-Based Frequency Combs

The area where these platforms have arguably had the greatest impact is in integrated OPOs based on ring resonators. These devices have significant potential for many applications including spectroscopy and metrology [69, 77, 78] as well as the ability to provide an on-chip link between the RF and the optical domains. It is not our intention to comprehensively review the field of micro-cavity based frequency combs, as this has recently been done [69], but to highlight the strengths and versatility specifically of these CMOS compatible platforms in producing frequency combs.

Micro-cavities effectively enhance nonlinear optical processes, particularly FWM involving a continuous-wave (CW) pump, signal and idler beams with frequencies ($\omega_{Pump}$, $\omega_{Idler}$, $\omega_{Signal}$) related by energy conservation: $\omega_{Idler} = 2\omega_{Pump} - \omega_{Signal}$. This process can occur either classically (with a separate input signal at $\omega_{Signal}$) or spontaneously (without one) – the former being the basis for many all-optical signal processing functionalities. Very low power operation was first demonstrated in silica and single crystal CaF and MgF micro-toroids and spheres with Q-factors ranging from $10^7$ to $10^{10}$ [69, 77, 78]. Achieving phase-matching of the propagation constants for three interacting waves is essential for efficient FWM, which for the microcavities is equivalent to having near-equal resonance spacings, or a constant FSR, with due allowance for the Kerr-induced resonance shifts [57]. This results in the pump, signal and idler waves all being in resonance – a triple resonance that greatly reduces the power requirement for the round-trip parametric gain to exceed the loss, thus producing oscillation. Phase matching, achieved by obtaining low and anomalous waveguide dispersion can be realised with a suitable design of the waveguide cross-section. Once oscillation is achieved via pure spontaneous (degenerate) FWM (with only a pump present), "cascaded" FWM among different cavity modes takes over, resulting in the generation of a frequency comb of precisely spaced modes in the frequency domain. However, the enhancement of the fields in the cavity (all fields in the case of phase matching) that is responsible for lowering the oscillation threshold, also enhances the nonlinear losses, and it is primarily for this reason that oscillation in silicon ring resonators in the telecom band where the FOM < 1, has not been achieved.

As a prelude to achieving parametric oscillation in ring resonators, low power CW nonlinear optics (FWM) was demonstrated for moderate Q-factor (65,000) Hydex [59] ring resonators using only a few milliwatts of CW pump power. This yielded an idler that was almost exactly on resonance indicating that the dispersion was indeed negligible. Although the $\gamma$ factor was much lower than for SOI ring resonators [79], the negligible nonlinear absorption allows the use of higher pump powers (+24dBm) where high bit rate all-optical signal processing is typically performed [80 - 82].

In early 2010 OPOs were simultaneously reported in SiN [57] and Hydex [58] CMOS-compatible integration platforms with ring resonators having much lower Q factors than previous micro-cavity oscillators. Hence these devices had much less sensitivity to environmental perturbations and avoided the need for delicate tapered fibre coupling.

Figure 2a shows a Hydex four-port micro-ring resonator with a Q factor of 1.2 million along with the corresponding optical transmission spectrum. The optical frequency comb generated by this Hydex device [58] exhibited a very wide spacing of almost 60 nm when pumped at 1,544.15 nm, in the anomalous GVD regime (Figure 2b). The output power versus pump power showed a significantly high single line differential slope efficiency above threshold of 7.4%, with a maximum power of 9 mW achieved in all oscillating modes out of both ports, representing a remarkable absolute total conversion efficiency of 9%. When pumping at a slightly different wavelength closer to the zero-GVD wavelength (but still in the anomalous regime), the device oscillated with significantly different spacing of 28.15 nm. These observations are consistent with parametric gain based on a combination of FWM and the parametric, or modulational instability (MI), gain described above, where the spacing depends on the waveguide dispersion characteristics and agrees well with calculations. This illustrates the degree of freedom one can achieve in varying the frequency comb spacing – chiefly through dispersion engineering, and so the nonlinear response is not restricted by the FSR of the resonator itself. The trade-off is that MI generated combs can themselves further seed sub-combs that are poorly related in terms of coherence to the original comb, limiting the degree to which modelocking, or ultrashort pulse generation, can be achieved [65, 68] (See next section).

Figure 2c shows a 2-port SiN micro-ring resonator [57] with a 58 μm radius resonator (Q factor = 500,000, FSR = 403GHz) with dimensions designed to yield anomalous GVD in the C-band with a zero GVD point at 1,610 nm. Optical parametric oscillation was first reported in SiN [57] by resonantly pumping the rings with CW light near 1550 nm using a soft 'thermal lock' process in which the cavity heating is counteracted by diffusive cooling. Oscillation of multiple lines over a very broad (>200 nm) wavelength range was achieved (Figure 2d) at a pump threshold of 50 mW. Eighty-seven new frequencies were generated between 1,450 and 1,750 nm, corresponding to wavelengths covering the S, C, L and U communications bands. Several designs were employed with different ring radii, or FSR. A smaller ring with a Q factor of 100,000 generated oscillation in 20 resonator modes with THz mode spacing when pumped with modest input powers (150 mW). The observation that these devices oscillated with a frequency spacing equal to the FSR of the resonator, in contrast with the Hydex device that oscillated with a frequency spacing given by the MI gain peak, is perhaps better understood in the light of recent studies (see discussion below) on comb formation and dynamics, and is a result of the subtle interplay of the waveguide dispersion and precise power and detuning of the pump wavelength.

## Advanced Frequency Combs

Since these initial demonstrations of multiple wavelength oscillation, significant progress has been made in advanced comb generation, including both very wide bandwidth octave spanning combs [66] and very low (sub 100GHz) FSR spacing combs[67].

The development of microresonator-based frequency combs with a free spectral range (FSR) significantly less than 100 GHz is critical to provide a direct link between the optical and electrical domains in order to produce highly stable microwave signals [67, 83 - 85] detectable with photodiodes. The challenge is that simple ring geometries with sub-100GHz FSR spacings do not fit on typical single e-beam fields and so novel ring geometries such as spirals need to be employed (Fig 3a-c). Figure 3 shows spiral ring resonators with unique geometries for different FSRs below 100GHz [67], all having a constant semicircular coupling region to enable critical coupling between the bus and resonator, independent of path length. Bends in the resonators had radii > 100 μm to ensure that bend induced dispersion was negligible, a critical requirement for proper operation of the frequency comb. The experimental spectra for 80, 40, and 20 GHz combs are shown in Figure 3d-f, typically requiring about 2W pump power to fill the entire comb spans.

Octave-spanning frequency combs are of great interest for spectroscopy, precision frequency metrology, and optical clocks and are highly desirable for comb self-stabilization using *f* to 2*f* interferometry for precision measurement of absolute optical frequencies [85-87]. Figure 3g shows an optical frequency comb in a SiN ring resonator spanning more than an octave [66] from 1170 to 2350 nm, corresponding to 128 THz. This is achieved by suitable dispersion engineering and employing higher pump powers of up to 400 mW inside the waveguide, detuned slightly from a cavity resonance. Figure 3h shows the simulated dispersion for nanowires with varying widths (1200, 1650, and 2000 nm) indicating that large anomalous-GVD bandwidths spanning nearly an octave are possible with appropriate design. These results represent a significant step toward a stabilized, robust integrated frequency comb source that can be scaled to other wavelengths.

## Supercontinuum Generation

Very broadband continuous spectra can be generated by injecting ultrashort modelocked pulse trains into suitably designed waveguides. These super-continuum (SCG) spectra are of interest for similar reasons to octave-spanning frequency combs. Wide bandwidth SCG has been demonstrated in microstructured fibers [88], chalcogenide glass waveguides [89], periodically poled lithium niobate (PPLN) [90], and in Si beyond 2 µm [91]. Hydex and SiN offer the advantage of much lower linear and nonlinear losses as well as transparency well into the visible. It is interesting to compare the broadband frequency combs with SCG in Hydex waveguides [92] where a spectral width > 350 nm was achieved (limited by experimental measurement capability), as well as with SCG in 1100-nm wide SiN nanowires [93]. When pumping at 1335 nm in this platform SCG resulted in a spectrum spanning 1.6 octaves, from 665 nm to 2025 nm. We note that the SiN results represent the broadest recorded SCG to date in a CMOS compatible chip. Remarkably, both of these results were enabled by a high effective nonlinearity, negligible TPA and most significantly, by very flexible dispersion engineering. SCG is significantly enhanced if the pulse is launched near a zero group-velocity dispersion (GVD) point or in the anomalous GVD regime [93-96]. The former minimizes temporal pulse broadening, thereby preserving high peak powers and thus maintaining a strong nonlinear interaction. The latter regime enables soliton propagation, whose dynamics can contribute to spectral broadening [97]. The higher flexibility of dispersion engineering in SiN as compared to Hydex is partly related to its higher available core/cladding index contrast and is probably one of its most significant advantages.

## Frequency Comb Dynamics and Coherence

As with conventional modelocked lasers, parametric frequency combs can also potentially serve as sources of ultrashort laser pulses from the visible to the mid-infrared at repetition rates in the GHz to THz regimes. The last two years have seen significant breakthroughs in understanding the dynamics and coherence behavior associated with frequency comb formation [65-70]. These reports have revealed complex and distinct paths to comb formation that can result in widely varying degrees of coherence, wavelength spacing, and RF stability of these sources. This field has recently been highlighted by the achievement of an ultrastable, ultrashort optical pulse source via modelocking [15] based on an integrated microcavity. Understanding and harnessing the coherence properties of these monolithic frequency comb sources is crucial for exploiting their full potential in the temporal domain, as well as for generally bringing this promising technology to practical fruition.

The first investigation of the coherence properties of on-chip oscillators [68] focused on phase tuning of the comb via programmable optical pulse shaping. This "line-by-line" pulse shaping of microresonator frequency combs was enabled by the relatively large mode spacings and represented a significant development in the field of optical arbitrary waveform generation [83 - 85]. In this approach, transform limited pulses can be realized for any spectral phase signature of coherent combs by appropriately compensating the relative phase of the different comb lines (Figure 4a-d). These time domain experiments revealed different pathways to comb formation in terms of phase coherence properties, where the ability to effectively modelock the combs varies significantly depending on many factors such as pump conditions and waveguide dispersion.

More recent studies on the dynamics of comb formation [65] (Figure 4e-f) have shown that the initial oscillation often begins with frequency spacings near the parametric gain peak, which can vary widely from as little as the FSR spacing itself to as wide as 100nm or more. Cascaded FWM then replicates a comb with this initial spacing. More complex dynamics arise when these comb lines themselves seed their own mini-comb bands based on the local dispersion and pumping conditions, often with a spacing at or near the cavity FSR in that wavelength region. While these "sub-combs" maintain coherence within themselves, they are not coherently related to the sub-combs generated by other lines of the initial, more widely spaced, comb, and so this results in a limited overall coherence. One approach to obtain high overall coherence is to achieve initial comb oscillation at the FSR. Figure 4 shows the output of a ring resonator under different pumping conditions [65] showing the transition from an incoherent state with well separated frequency sub-combs to a coherent state featured by a much improved ability to lock the modes and thus to produce optical pulses, thus resulting in much lower RF noise.

Further studies [70] of the optical and radio-frequency properties of parametric frequency combs in SiN microresonators show a transition to a modelocked state where ultrashort pulse generation is possible. From a 25-nm filtered section of the 300-nm comb spectrum, sub-200-fs pulses were observed with calculations indicating that the pulse generation process is consistent with soliton modelocking. This is consistent with very recent work involving comb generation in $MgF_2$ microresonators [98] and could be explained by the formation of temporal cavity solitons [15, 99], where nonlinearity and a coherent driving beam compensate contributions from both dispersion and loss.

## Ultrashort Pulse Sources

The first demonstration of stable modelocking of a microresonator frequency comb was recently achieved (Figure 5a) [15, 71] by embedding the resonator in an active fiber loop where the microresonator is used as both a linear filter and a nonlinear element. This scheme, termed Filter-Driven Four-Wave-Mixing (FD-FWM), is an advancement inspired by dissipative FWM (DFWM) [100-102] where the nonlinear interaction occurs in-fibre and is then "filtered" separately by a linear FP filter. This new approach is more efficient and so allows for substantially reduced main cavity lengths that lead in turn to highly stable operation − something that has so far eluded DFWM lasers. In DFWM lasers the main cavity mode spacing is much finer than the microcavity, allowing many cavity modes to oscillate within the filter resonance, an unfavourable condition which gives rise to "supermode instability". FD-FWM allows a substantial reduction in the main cavity length so that only a very small number of cavity modes exist within each micro-resonator resonance, allowing for the possibility of true single mode operation. FD-FWM has achieved highly stable operation at high repetition rates over a large range of operating conditions, robust to external (i.e., thermal) perturbations. It also yields much narrower linewidths than ultrashort cavity lasers because it has a much smaller Schawlow-Towns phase noise limit [102, 103].

Figure 5a shows the experimental configuration of the loop laser. It compares the time resolved optical waveforms (measured by autocorrelation, Fig. 5b,c), the RF temporal output (Fig. d,e), the radio-frequency (RF) spectra (Fig. 5f,g), and the optical spectra (Fig. 5d,e inset), of a short and long cavity length laser, clearly showing that the RF spectrum for the short cavity laser is highly stable whereas the long cavity length laser not only shows large, long timescale, instabilities in the RF output, but is also unstable on a short timescale reflected by the fact that the optical autocorrelation traces show a very limited contrast ratio – a hallmark of unstable oscillation [65, 68]. Recently [71] stable modelocked laser operation with two modes oscillating within each ring cavity was demonstrated, which produced a highly pure RF tone over the modelocked train, enabling the high resolution RF spectral measurement of the optical linewidth that displayed a remarkably narrow width of about 10kHz.

## Ultrafast Phase Sensitive Pulse Measurements

Coherent optical communications [18, 19] have created a compelling need for ultrafast phase-sensitive measurement techniques operating at milliwatt peak power levels. Ultrafast optical signal measurements have been achieved using time-lens temporal imaging on a silicon chip [105, 106] and waveguide-based Frequency-Resolved Optical Gating (FROG) [107], but these are either phase insensitive or require waveguide-based tunable delay lines - a difficult challenge. Recently [14], a device capable of both amplitude and phase characterization of ultrafast optical pulses was reported, operating via a novel variation of Spectral Phase Interferometry for Direct Electric-Field Reconstruction (SPIDER) [108 - 115] based on FWM in a Hydex waveguides. Here, pulse reconstruction was obtained with the aid of a synchronized incoherently related clock pulse.

SPIDER approaches are ultrafast, single-shot, and have a simple, direct and robust phase retrieval procedure. However they traditionally have employed either three-wave mixing (TWM) or linear electro-optic modulation, both of which require a second-order nonlinearity that is absent in CMOS compatible platforms. Also, typical SPIDER methods work best with optical pulses shorter than 100 fs [114] (small time-bandwidth products (TBP)) and peak powers > 10 kW, and hence are not ideally suited to telecommunications. The device reported in [14] measured pulses with peak powers below 100 mW, with a frequency bandwidth greater than a terahertz and on pulses as wide as 100 ps, yielding a record TBP > 100 [114]. The technique employed in this device is commonly referred to as X-SPIDER.

Figure 6a shows a schematic of the device. The pulse under test (PUT) is split into two replicas and nonlinearly mixed with a highly chirped pump inside the chip. The resulting output is then captured with a spectrometer and numerically processed to extract the complete information (amplitude and phase) of the incident pulse. Figure 6b-c compares the results of the X-SPIDER device using both a standard algorithm and a new extended (Fresnel) phase-recovery extraction algorithm [14, 116] designed for pulses with a large time bandwidth product (TBP), with the measures obtained for the same pulses from SHG-based FROG measurements. As expected, for low TBP pulses (short-pulse regime) the SPIDER device yielded identical results when executed with both algorithms, and also agreed with the FROG spectrogram. For large TBP pulses (highly chirped, long pulsewidths) the X-SPIDER results obtained with the new algorithm agreed very well with the FROG trace, whereas results obtained using the standard phase-recovery algorithm were unable to accurately reproduce the pulse.

# Future Challenges, Opportunities

The success of these platforms for nonlinear optics in the telecommunications band has been due to a favorable combination of very low linear loss, negligible TPA, a moderately high nonlinearity, together with the ability to engineer dispersion, their high quality growth and fabrication processes, and their excellent material reliability. However, the *ideal* platform would have a much larger nonlinearity - comparable to, or even larger than, silicon – while at the same time displaying a much larger FOM than silicon.

Table I compares the nonlinear parameters of key CMOS compatible platforms including crystalline and amorphous silicon, SiN, Hydex and fused silica. It is remarkable that SiN and Hydex perform so well given that their $n_2$ is typically only 5 to 10 times larger than that of fused silica and their $\gamma$ factors range from 200 (SiN) to 1000 times less than what is typical for SOI nanowires. This fact really highlights how critical low linear, and particularly nonlinear, losses are in all-optical platforms.

Amorphous silicon, studied as a nonlinear material [117] and platform for linear photonics [118, 119] for some time, was recently proposed [120] as an alternative to SOI for nonlinear optics in the telecom band. Although initial measurements yielded a FOM no better than c-Si (~0.5) [120, 121], more recent results have shown FOMs ranging from 1 [122] to as high as 2 [123, 124], allowing very high parametric gain (+26dB) over the C-band [125]. While a key problem for this material has been a lack of stability [126], very recently a-Si nanowires were demonstrated [127] that displayed a combination of high FOM of 5, high $n_2$ (3-4 times that of crystalline silicon) *and* good material stability at telecom wavelengths. A key goal of all optical chips is to reduce both device footprint and operating power. The dramatic improvement in both FOM and nonlinearity of a-Si raises the possibility of using slow-light structures [128, 129] to allow devices to operate at mW power levels with sub-millimeter lengths, and so a-Si remains an extremely promising platform for future all-optical chips.

## Conclusions

We have reviewed the recent progress in two important CMOS compatible glasses that are potential alternatives to SOI for integrated optical platforms for nonlinear optics – silicon nitride and Hydex glass. These new platforms have enabled the realization of devices possessing many novel all optical functions at telecom wavelengths. The combination of negligible nonlinear (two-photon) absorption, low linear loss, the ability to engineer dispersion, and the moderately high nonlinearities has enabled these platforms to achieve new capabilities not possible in c-Si because of its low FOM. These platforms will likely have a significant impact on future all-optical devices, complimenting the substantial achievements already made in silicon nanophotonics. Their high performance, reliability and manufacturability, combined with intrinsic compatibility with electronic-chip manufacturing (CMOS), raises the prospect of practical platforms for future low-cost nonlinear all-optical PICs that offer a smaller footprint, lower energy consumption and lower cost than present solutions.

# Table I

Nonlinear parameters for CMOS compatible optical platforms

|  | a-Si [127] | c-Si [2, 47] | SiN [55-57] | Hydex [63, 76] |
|---|---|---|---|---|
| $n_2$ (x fused silica[1]) | 700 | 175 | 10 | 5 |
| $\gamma$ [W$^{-1}$m$^{-1}$] | 1200 | 300 | 1.4 | 0.25 |
| $\beta_{TPA}$ [cm/GW] | 0.25 | 0.9 | negligible[2] | negligible[3] |
| **FOM** | 5 | 0.3 | >> 1 | >> 1 |

[1] $n_2$ for fused silica = 2.6 x 10$^{-20}$ m$^2$/W [1]

[2] no nonlinear absorption has been observed in SiN nanowires.

[3] no nonlinear absorption has been observed in Hydex waveguides up to intensities of 25GW/cm$^2$ [63].

## Acknowledgements

The authors would like to acknowledge financial support from the Australian Research Council Discovery Project and Centres of Excellence programs, the Canadian Natural Sciences and Engineering Research Council (NSERC), the Defense Advanced Research Projects Agency (DARPA) and the National Science Foundation. We would also like to thank A. Pasquazi and M. Peccianti for proof-reading the manuscript.

Correspondence and requests for materials should be addressed to D.Moss (david.moss@rmit.edu.au)

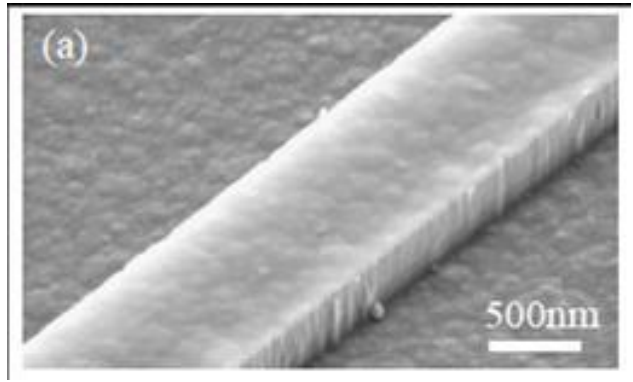

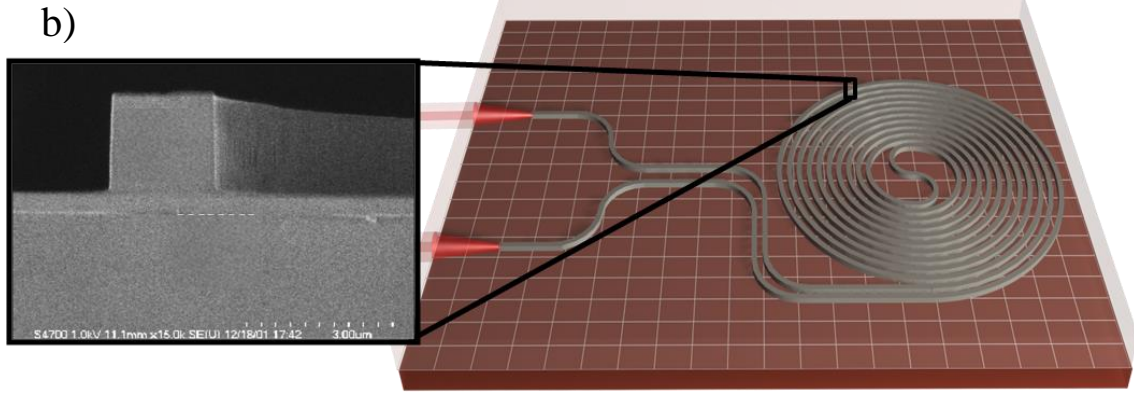

Figure 1. Silicon nitride nanowires (a) and Hydex (b) waveguides. (a) SEM micrograph of silicon nitride nanowires reported in [55, 56]. The nanowire dimensions are 500nm thick by 1 μm wide.

(b) Schematic and SEM image of the cross-section of a 45cm long spiral Hydex glass waveguide [14] prior to the final deposition of the SiO2 upper cladding. The waveguide core is 1.45 μm  x 1.5 μm and is low-loss, high-index (n = 1.7) doped silica glass and is buried within a SiO2 cladding (core–cladding contrast, 17%).

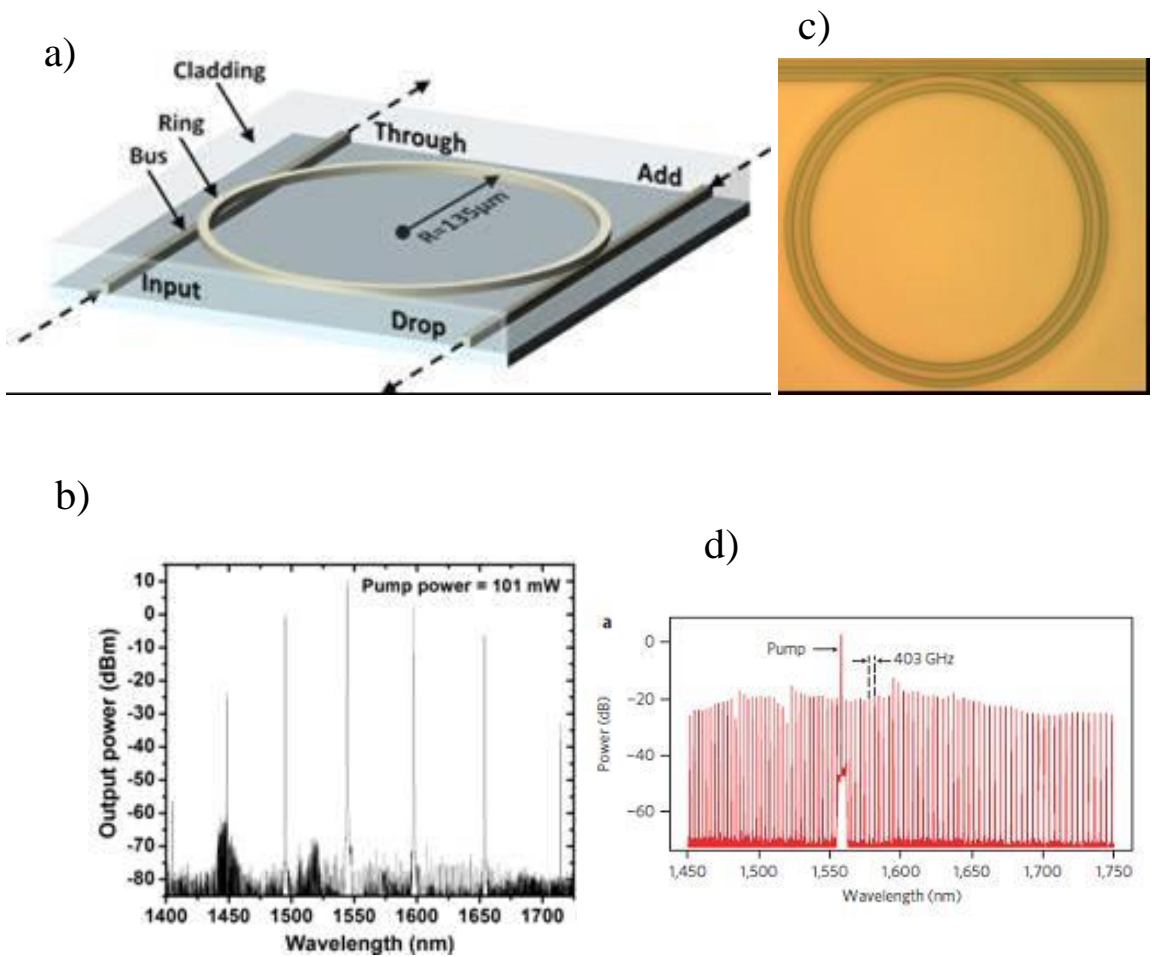

Figure 2. Integrated OPO multiple wavelength sources in Hydex (a-d) [56] and SiN [55] (f-h) ring resonators.

(a) Hydex four-port microring resonator [55] (fibre pigtails not shown) with Q factor of 1.2x10$^6$ and diameter of 270µm;
(b) Output spectra of an Hydex hyper-parametric oscillator for a pump power of 101mW injected on resonance at 1,544.15 nm (TM polarization).
(c) SEM image of a SiN microring resonator [57] (58µm radius and Q= 500,000, FSR= 403 GHz) coupled to a bus waveguide, with a cross section height of 711 nm, a base width of 1,700 nm and a sidewall angle of 20 degrees, giving anomalous GVD in the C-band and a zero-GVD point at 1,610 nm;
(d) Output spectra of a 58-µm-radius SiN ring resonator OPO with a single pump wavelength tuned to a resonance at 1,557.8 nm, showing numerous narrow linewidths at precisely defined wavelengths [55]. The 87 generated wavelengths were equally spaced in frequency, with a FSR of 3.2 nm.

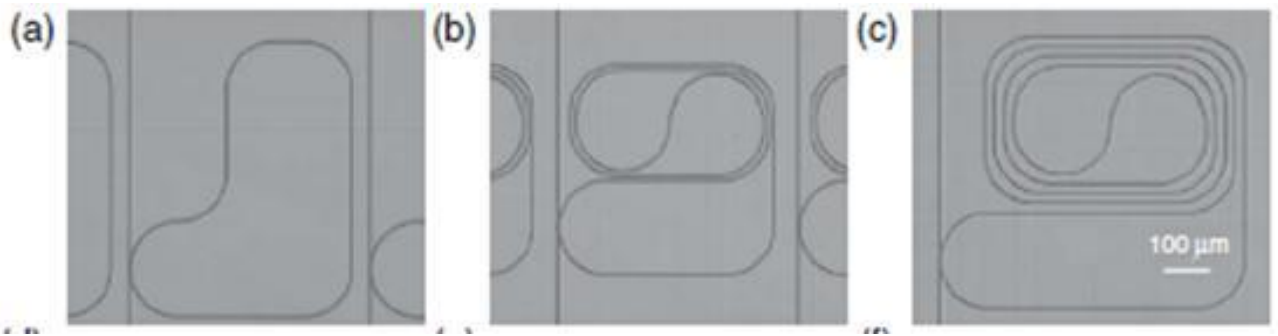
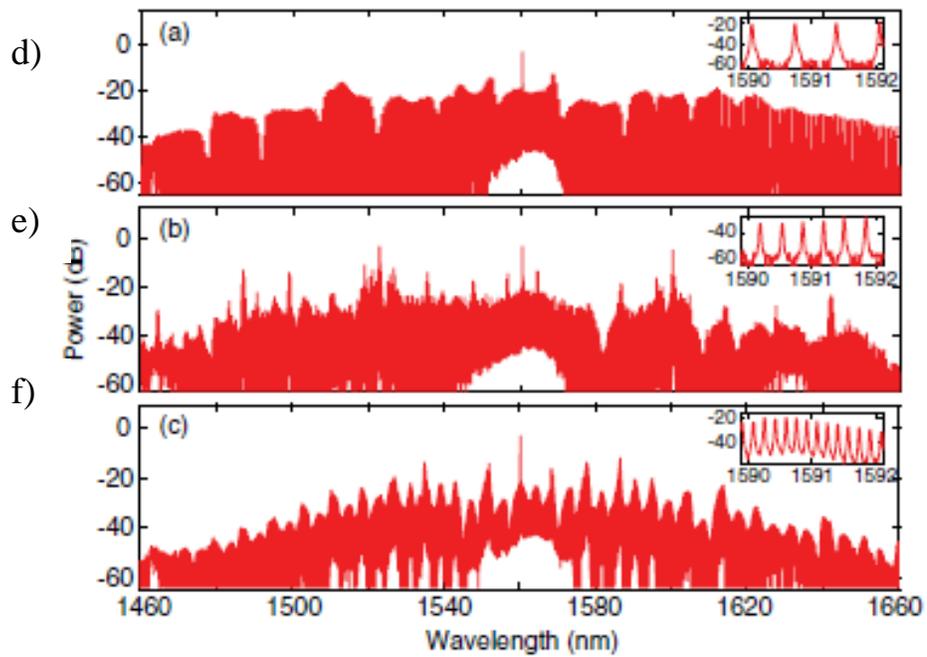
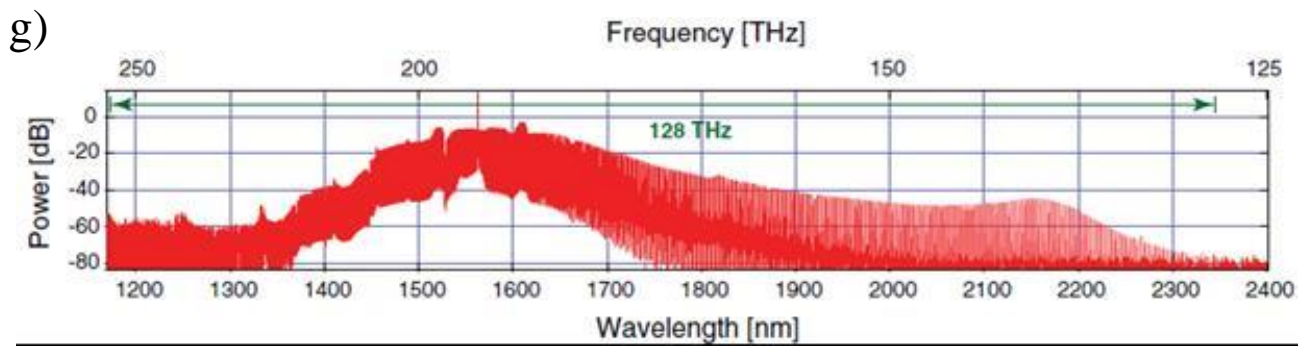

h)

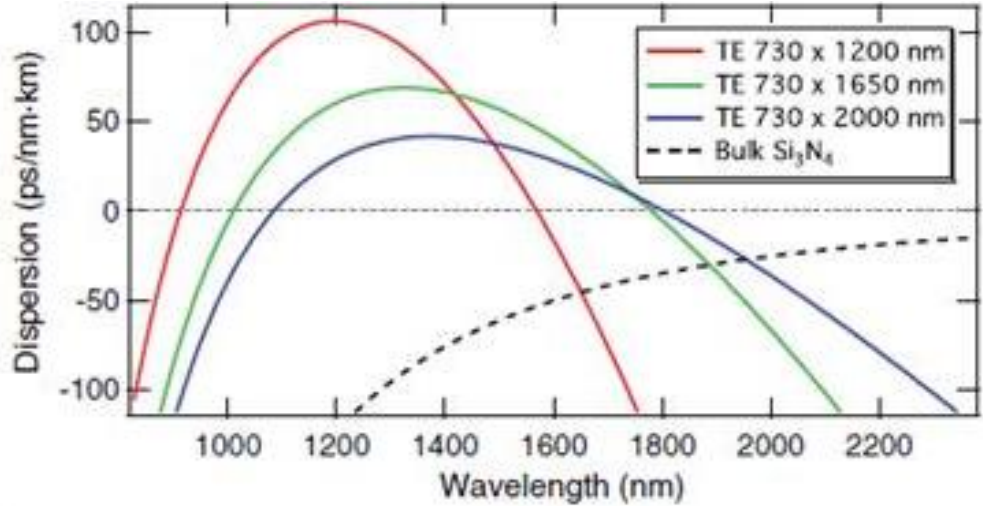

**Figure 3. Advanced frequency combs in SiN ring resonators.** (a-f) Sub-100GHz spacing SiN ring resonators [67]. Micrographs of the (a) 80GHz, (b) 40GHz, and (c) 20 GHz FSR resonators and the corresponding linear transmission spectra. The nanowire cross sections were 725nm by 1650 nm, with a minimum microring bend diameter of 200 μm and a loaded Q of 100,000. Output spectra are 300 nm wide for the 80GHz and 40GHz FSR rings and 200 nm for the 20GHz FSR ring, respectively. A 2 nm section of each comb is given as an inset in each figure to illustrate the spacing of the comb lines.

(g) Optical spectrum of octave-spanning parametric frequency comb generated in a SiN ring resonator [66] (h) dispersion simulations for the fundamental TE mode of a SiN waveguide with a height of 730nm and widths of 1200, 1650, and 2000 nm, respectively. The dashed curve shows the dispersion for bulk silicon nitride. a) to f) ref [67], g)-h) ref [66].

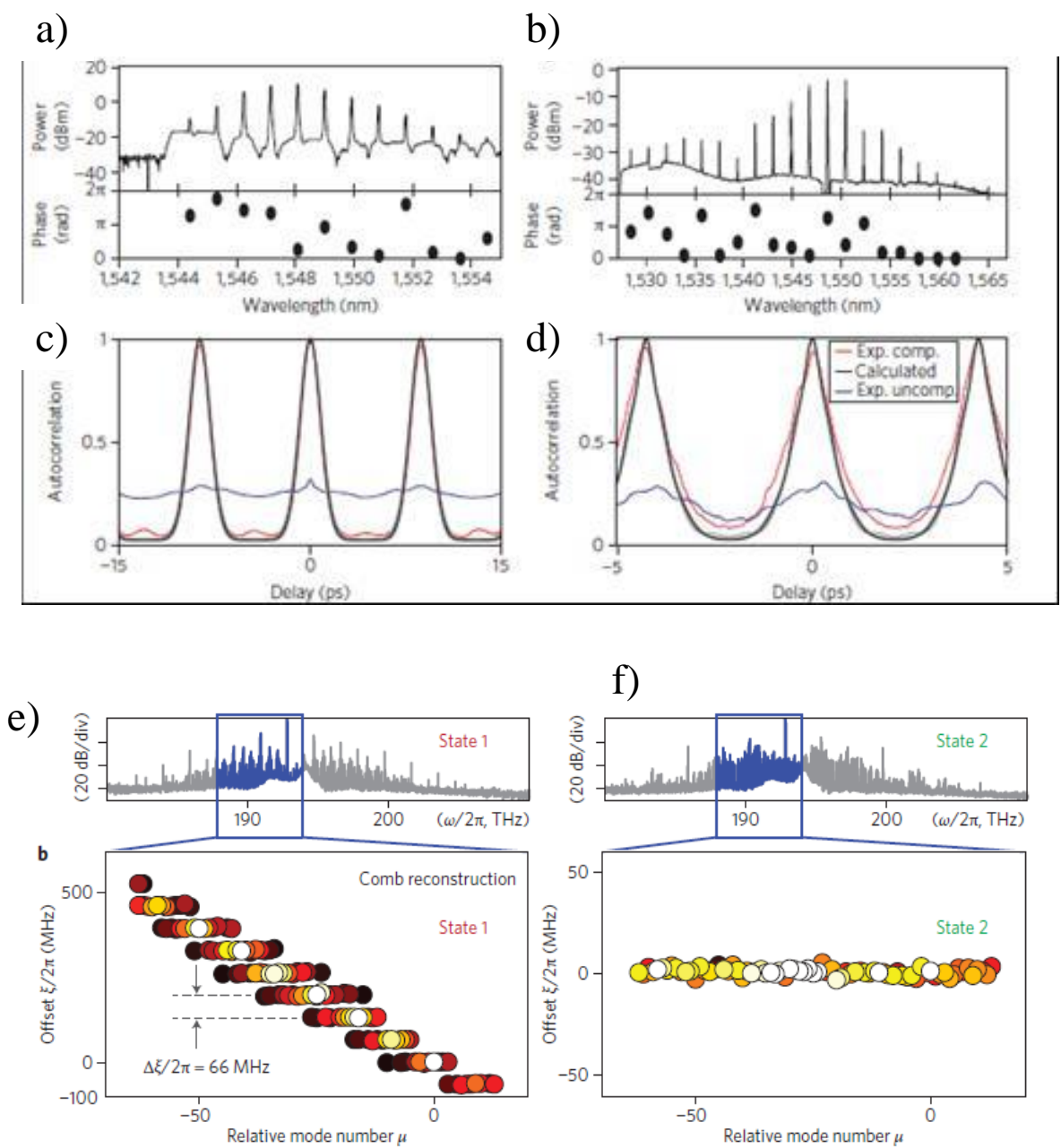

**Figure 4 Coherence, frequency comb formation dynamics, and ultrashort pulse modelocked sources.**

(a-d). Study of Frequency comb coherence properties from [68]. (a) The output spectra generated by a high-Q silicon nitride microring shows the ability of type-I Kerr combs to perform pulse compression . (a,b) Spectra of the combs after processing with a pulse shaper, together with the phase applied to the liquid-crystal modulator pixels a required to achieve optimum SHG signals.

(c,d) Autocorrelation traces corresponding to (a,b). The red lines depict the compressed pulses after a phase correction is applied, the blue lines are uncompressed pulses and the black lines are calculated by taking the spectra shown in a and b and assuming a flat spectral phase. The contrast ratios of the autocorrelations measured after phase compensation are 14:1 (c) and 12:1 (d), respectively. Light grey traces show the range of the simulated autocorrelation traces.

(e - f) Studies of coherence evolution in SiN microring resonators from [65] showing the optical spectra for a transition from a low phase noise Kerr comb (e) – state 1 - to a high coherence output (f) - state 2 for the two microresonator comb states (pump power 6W). State 2 evolves from state 1 when reducing the detuning of the pump laser. A transition is observed from multiple sub-combs to a single sub-comb over the bandwidth of the Kerr comb reconstruction. In state 1, all sub-combs have the same mode spacing, but have different offsets that differ by a constant value of 66 MHz. In the transition from state 1 to state 2, the amplitude noise peak resulting from the beating between overlapping offset sub-combs disappears [65].

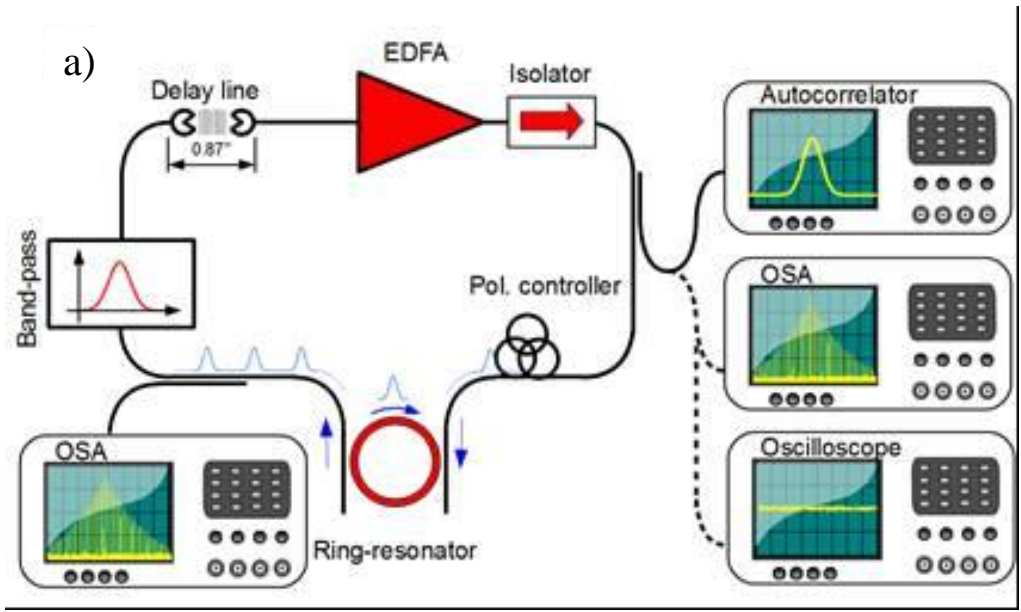

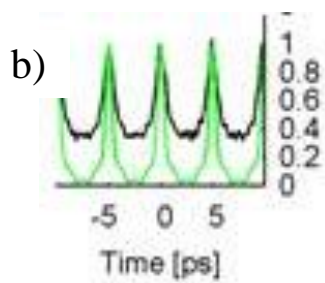
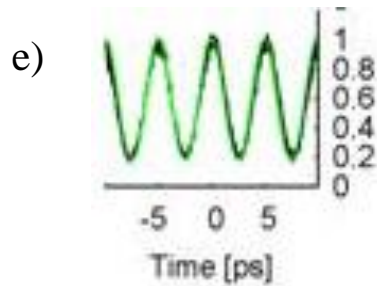

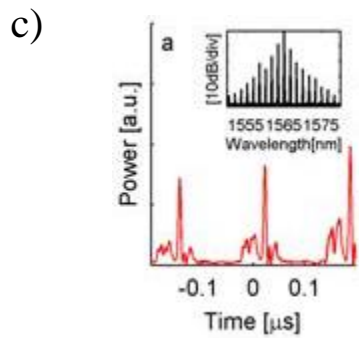
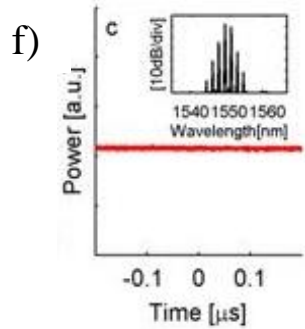

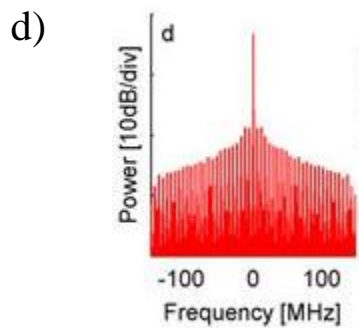
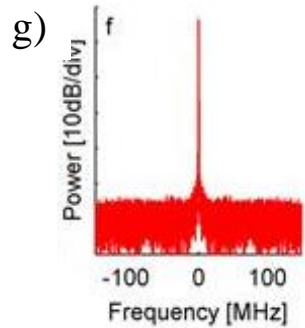

**Figure 5. Microresonator based modelocked fiber loop laser** [15]. (a) Experimental configuration for a fiber loop modelocked laser based on filter driven four wave mixing (FDFWM) in a microring resonator, where the resonator performs the dual function of both linear filtering and nonlinear interaction.

(b - d) Results for the unstable operation regime with a long main cavity length (L=33m, FSR = 6.0 MHz);

(e - g) Results for the stable operation regime for a short cavity main length (L=3m, FSR=68.5MHz) laser. Stable oscillation was obtained by adjusting the phase of the cavity modes for the short main cavity length laser relative to the ring resonator modes via a free space delay line.

(b, e) Experimental optical temporal waveforms measured via optical autocorrelation for unstable (b) and stable (e) operation. Autocorrelation plots show theoretical calculations (green) starting from the experimental optical spectra for a fully coherent and transform-limited system calculated by considering each line of the experimental optical spectra as being perfectly monochromatic and in-phase with the others, yielding an output pulse with a FWHM of 730 fs for the highest excitation power condition. The measured autocorrelation for the long (unstable) cavity laser (b) shows a considerably higher background than the expected autocorrelation (50:1). Conversely, the calculated autocorrelation trace for the short-cavity (stable) laser (e) perfectly matches the measured trace, suggesting stable modelocking, and corresponding to a transform-limited pulse width (FWHM) of 2.3 ps with a peak to background ratio of 5:1.

(c, f) Radio frequency (RF) temporal output for the unstable (c) and the stable (f) configurations. (d, g) RF spectra for the unstable (d) and the stable (g) configurations.

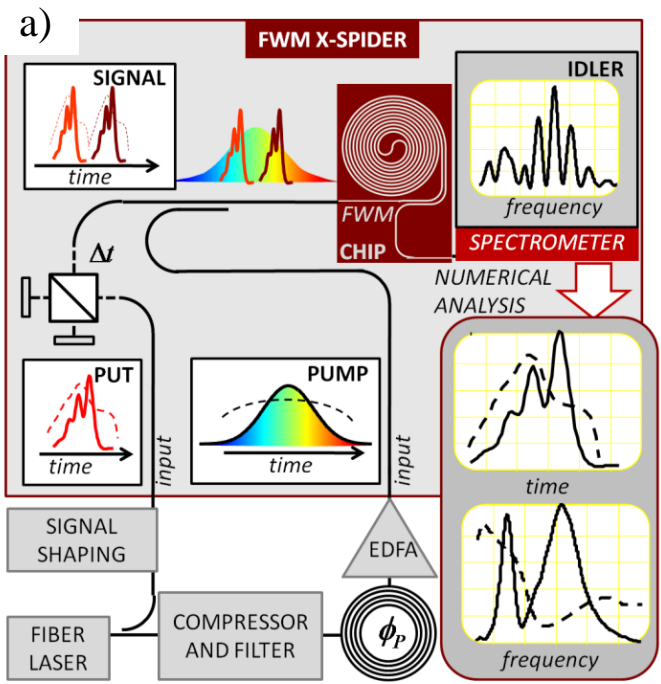

b)
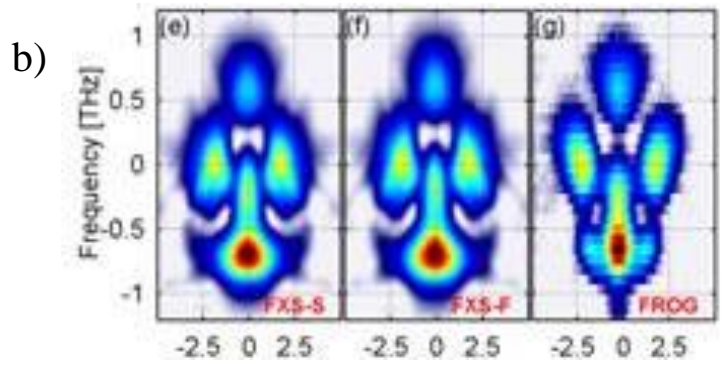

c)
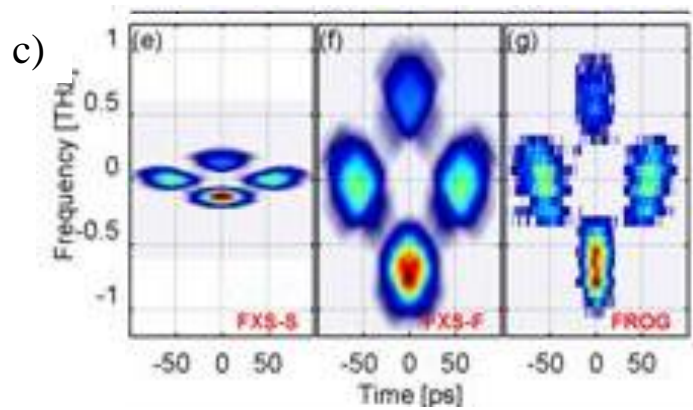

**Figure 6.** Phase and amplitude measurement of ultrafast optical pulses using spectral phase interferometry for direct electric field reconstruction (SPIDER) [14].

(a) X-SPIDER based on FWM in a Hydex spiral waveguide. The PUT (pulse under test) is split in two replicas and nonlinearly mixed with a highly chirped pump inside the chip (extracted from the same laser source in these experiments). The resulting output is captured with a spectrometer and numerically processed with a suitable algorithm to extract the complete information (amplitude and phase) of the incident pulse.

(b) Retrieved phase and amplitude profiles for pulses with time bandwidth products (TBP) of 5.

(c) Retrieved phase and amplitude profiles for pulses with time bandwidth products (TBP) of 100.

In (b) and (c) the left hand plots were obtained by the X-SPIDER device using a standard algorithm while the middle plots used a new algorithm [14, 116] suitable for large time-bandwidth product (TBP) pulses. The plots at the right are the experimentally measured FROG SHG measurements.